\documentclass[letterpaper, 12pt]{article}[2000/05/19]
\usepackage[english]{babel}
\usepackage{amsfonts,amsmath,amssymb,amsthm,latexsym,amscd,mathrsfs}
\usepackage{ifthen,cite}
\usepackage[bookmarksnumbered=true]{hyperref}

\hypersetup{pdfpagetransition={Split}}

\newcommand{\evenhead}{Author \ name}
\newcommand{\oddhead}{Article \ name}
\newcommand{\theArticleName}{Article \ name}

\newcommand{\FirstPageHeading}[1]{\thispagestyle{empty}%
\noindent\raisebox{0pt}[0pt][0pt]{\makebox[\textwidth]{\protect\footnotesize \sf }}\par}

\newcommand{\ArticleName}[1]{\renewcommand{\theArticleName}{#1}\vspace{-2mm}\par\noindent {\LARGE\bf  #1\par}}
\newcommand{\Author}[1]{\vspace{5mm}\par\noindent {\Large  #1\par} \par\vspace{2mm}\par}
\newcommand{\Address}[1]{\vspace{2mm}\par\noindent {\it #1} \par}
\newcommand{\Email}[1]{\ifthenelse{\equal{#1}{}}{}{\par\noindent {\rm E-mail: }{\it  #1} \par}}
\newcommand{\URLaddress}[1]{\ifthenelse{\equal{#1}{}}{}{\par\noindent {\rm URL: }{\tt  #1} \par}}
\newcommand{\EmailD}[1]{\ifthenelse{\equal{#1}{}}{}{\par\noindent {$\phantom{\dag}$~\rm E-mail: }{\it  #1} \par}}
\newcommand{\URLaddressD}[1]{\ifthenelse{\equal{#1}{}}{}{\par\noindent {$\phantom{\dag}$~\rm URL: }{\tt  #1} \par}}

\newcommand{\Abstract}[1]{\vspace{6mm}\par\noindent\hspace*{10mm}
\parbox{140mm}{\small {\bf Abstract.} #1}\par}
\newcommand{\Keywords}[1]{\vspace{3mm}\par\noindent\hspace*{10mm}
\parbox{140mm}{\small {\bf Key words:} \rm #1}\par}
\newcommand{\Classification}[1]{\vspace{3mm}\par\noindent\hspace*{10mm}
\parbox{140mm}{\small {\it 2000 Mathematics Subject Classification:} \rm #1}\vspace{3mm}\par}
\newcommand{\ShortArticleName}[1]{\renewcommand{\oddhead}{#1}}
\newcommand{\AuthorNameForHeading}[1]{\renewcommand{\evenhead}{#1}}

\long\def\@makecaption#1#2{%\vskip\abovecaptionskip
  \sbox\@tempboxa{\small \textbf{#1.}\ \ #2}%
  \ifdim \wd\@tempboxa >\hsize
    {\small \textbf{#1.}\ \ #2}\par \else
    \global \@minipagefalse
    \hb@xt@\hsize{\hfil\box\@tempboxa\hfil}%
  \fi \vskip\belowcaptionskip}
\def\numberwithin#1#2{\@ifundefined{c@#1}{\@nocounterr{#1}}{%
  \@ifundefined{c@#2}{\@nocnterr{#2}}{%
  \@addtoreset{#1}{#2}%
  \toks@\@xp\@xp\@xp{\csname the#1\endcsname}%
  \@xp\xdef\csname the#1\endcsname
    {\@xp\@nx\csname the#2\endcsname.\the\toks@}}}}
\def\E^#1{{\buildrel #1 \over\vee}}

{\theoremstyle{definition}

}

\begin{document}

\FirstPageHeading{Yu.Yu. Fedchun and V.I. Gerasimenko}

\ShortArticleName{Kinetic equations in mathematical biology}

\AuthorNameForHeading{Yu.Yu. Fedchun and V.I. Gerasimenko}

\ArticleName{On Kinetic Equations Modeling Evolution of\\ Systems
            in Mathematical Biology}

\Author{Yu.Yu. Fedchun$^\ast$\footnote{E-mail:\hspace*{1mm}\emph{fedchun.yu@ukr.net}} and
        V.I. Gerasimenko$^\ast$$^\ast$\footnote{E-mail:\hspace*{1mm}\emph{gerasym@imath.kiev.ua}}}

\Address{$^\ast$\hspace*{1mm}Taras Shevchenko National University of Kyiv,\\
    \hspace*{4mm}Department of Mechanics and Mathematics,\\
    \hspace*{4mm}2, Academician Glushkov Av.,\\
    \hspace*{4mm}03187 Kyiv, Ukraine}

\Address{$^\ast$$^\ast$Institute of Mathematics of NAS of Ukraine,\\
         \hspace*{3mm}3, Tereshchenkivs'ka Str.,\\
         \hspace*{3mm}01601 Kyiv, Ukraine}

\bigskip

\Abstract{We develop a rigorous formalism for the description of the kinetic evolution of interacting
entities modeling systems in mathematical biology within the framework of the evolution of marginal
observables. For this purpose we construct the mean field asymptotic behavior of a solution of the
Cauchy problem of the dual BBGKY hierarchy (Bogolyubov-Born-Green-Kirkwood-Yvon) for marginal
observables of the dynamical systems based on the Markov jump processes, exhibiting the intrinsic
properties of the living entities. The constructed scaling limit is governed by the set of recurrence
evolution equations, namely by the dual Vlasov-type hierarchy. Moreover, the relationships of the
dual Vlasov hierarchy for the limit marginal observables with the Vlasov-type kinetic equation is
established.}

\bigskip

\vspace{2pc}
\Classification{35Q92; 37N25; 82C22.}

\makeatletter
\renewcommand{\@evenhead}{
\hspace*{-3pt}\raisebox{-7pt}[\headheight][0pt]{\vbox{\hbox to \textwidth {\thepage \hfil \evenhead}\vskip4pt \hrule}}}
\renewcommand{\@oddhead}{
\hspace*{-3pt}\raisebox{-7pt}[\headheight][0pt]{\vbox{\hbox to \textwidth {\oddhead \hfil \thepage}\vskip4pt\hrule}}}
\renewcommand{\@evenfoot}{}
\renewcommand{\@oddfoot}{}
\makeatother

\newpage
\vphantom{math}

\protect\tableofcontents
%\newpage
\vspace{0.5cm}
%\title{Title\tnoteref{label1}}

\section{Introduction}

Recently the considerable advance in solving the problem of rigorous modeling of the kinetic
evolution of systems with a large number of constituents (entities) of mathematical biology,
in particular, systems of large number of cells, is observed \cite{L08}-\cite{CDW} (and see
references cited therein).

The many-entity biological systems \cite{ALBA},\cite{L11} are dynamical systems displaying
a collective behavior which differs from the statistical behavior of usual gases \cite{CIP},
\cite{CGP97}. In the first place their own distinctive features is connected with the fact
that entities show the ability to retain various complexity features \cite{ALBA}-\cite{MJRLPRAS}.
To specify such nature of entities we consider the dynamical system suggested in \cite{L11},
\cite{LP} which is based on the Markov jump processes that must represent the intrinsic
properties of living creatures.

In the work we develop a new approach to the description of the kinetic evolution of large
number of interacting entities within the framework of the evolution of marginal observables.
To this end we describe the evolution of interacting entities by the dual BBGKY hierarchy for
marginal observables of such system and construct the mean field scaling asymptotics of its
solution. For obvious reasons the description of collective behavior in terms of observables
of living constituents of biological systems is more suitable in comparison with the formalism
of states.

We outline the structure of the paper and the main results. In introductory section 2 we adduce
the basic facts on the description of the evolution of systems of finitely many entities of
various subpopulations of mathematical biology introduced in paper \cite{L11}. In section 3 we
develop one more approach to the description of the evolution of many-entity systems in terms
of the hierarchies of evolution equations for marginal observables and marginal distribution
functions which underlie of kinetic models. In particular, a nonperturbative solution of the
Cauchy problem of the dual BBGKY hierarchy for the marginal observables is constructed. In
section 4 we prove the main result concerning to the description of the kinetic evolution of
interacting entities within the framework of the evolution of marginal observables. The mean
field asymptotics of a nonperturbative solution of the Cauchy problem of the dual BBGKY hierarchy
for entities is constructed. Furthermore, the relationships of the dual Vlasov hierarchy for the
limit marginal observables with the Vlasov-type kinetic equation is established.
Finally in section 5 we conclude with some observations and perspectives for future research.

\section{Preliminary facts}
A description of many-constituent systems is formulated in terms of two sets of objects: observables
and states. The functional of the mean value of observables defines a duality between observables
and states and as a consequence there exist two approaches to the description of the evolution
of such systems, namely in terms of the evolution equations for observables and for states. In
this section we adduce some preliminary facts about dynamics of finitely many entities of various
subpopulations described within the framework of nonequilibrium grand canonical ensemble \cite{CGP97}.

\subsection{The stochastic dynamics of many-entity systems}
We consider a system of entities of various $M$ subpopulations introduced in paper \cite{L11}
in case of non-fixed, i.e. arbitrary, but finite average number of entities. Every $i$th
entity is characterized by: $\textbf{u}_i=(j_i,u_i)\in\mathcal{J}\times\mathcal{U}$, where
$j_i\in\mathcal{J}\equiv(1,\ldots,M)$ is a number of its subpopulation, and
$u_i\in\mathcal{U}\subset\mathbb{R}^{d}$ is its microscopic state \cite{L11}.
The stochastic dynamics of entities of various subpopulations is described by the semigroup
$e^{t\Lambda}=\oplus_{n=0}^\infty e^{t\Lambda_n}$ of the Markov jump process defined on the 
space $C_\gamma$ of sequences  $b=(b_0,b_1,\ldots,b_n,\ldots)$ of measurable bounded 
functions $b_n(\textbf{u}_1,\ldots,\textbf{u}_n)$ that are symmetric with respect to 
permutations of the arguments $\textbf{u}_1,\ldots,\textbf{u}_n$ and equipped with the norm:
\begin{eqnarray*}
    &&\|b\|_{C_\gamma}=\max_{n\geq0}\frac{\gamma^n}{n!}\|b_n\|_{C_n}=
      \max_{n\geq0}\frac{\gamma^n}{n!}\max_{j_1,\ldots,j_n}\max_{u_1,\ldots,u_n}
      \big|b_n(\textbf{u}_1,\ldots,\textbf{u}_n)\big|,
\end{eqnarray*}
where $\gamma<1$ is a parameter. The operator $\Lambda_n$ (the Liouville operator of $n$ entities) 
is defined on the subspace $C_n$ of the space $C_\gamma$ and it has the following structure \cite{L11}:
\begin{eqnarray}\label{gen_obs_gener}
    &&\hskip-5mm(\Lambda_n b_n)(\textbf{u}_1,\ldots,\textbf{u}_n)\doteq
        \sum_{m=1}^M \varepsilon^{m-1}\sum_{i_1\neq\ldots\neq i_m=1}^n
        (\Lambda^{[m]}(i_1,\ldots,i_m)b_n)(\textbf{u}_1,\ldots,\textbf{u}_n)=\\
    &&\hskip-5mm=\sum_{m=1}^M \varepsilon^{m-1}\sum_{i_1\neq\ldots\neq i_m=1}^n
        a^{[m]}(\textbf{u}_{i_1},\ldots,\textbf{u}_{i_m})\big(\int_{\mathcal{J}\times\mathcal{U}}
        A^{[m]}(\textbf{v};\textbf{u}_{i_1},\ldots,\textbf{u}_{i_m})\times\nonumber\\
    &&\hskip+5mm\times b_n(\textbf{u}_1,\ldots,\textbf{u}_{i_1-1},\textbf{v},\textbf{u}_{i_1+1},
        \ldots\textbf{u}_n)d\textbf{v}-b_n(\textbf{u}_1,\ldots,\textbf{u}_n)\big)\nonumber,
\end{eqnarray}
where $\varepsilon>0$ is a scaling parameter \cite{Sp80}, the functions
$a^{[m]}(\textbf{u}_{i_1},\ldots,\textbf{u}_{i_m}),\,m\geq1,$ characterize the 
interaction between entities, in particular, in case of $m=1$ it is the interaction 
of entities with an external environment. These functions are measurable positive bounded 
functions on $(\mathcal{J}\times\mathcal{U})^n$ such that:
\begin{eqnarray*}
   &&0\leq a^{[m]}(\textbf{u}_{i_1},\ldots,\textbf{u}_{i_m})\leq a^{[m]}_*,
\end{eqnarray*}
where $a^{[m]}_*$ is some constant. The functions
$A^{[m]}(\textbf{v};\textbf{u}_{i_1},\ldots,\textbf{u}_{i_m}),\,m\geq1$, are measurable positive
integrable functions which describe the probability of the transition of the $i_1$ entity in the
microscopic state $u_{i_1}$ to the state $v$ as a result of the interaction with entities in the
states $u_{i_2},\ldots,u_{i_m}$ (in case of $m=1$ it is the interaction with an external environment).
The functions $A^{[m]}(\textbf{v};\textbf{u}_{i_1},\ldots,\textbf{u}_{i_m}),\,m\geq1$, satisfy
the conditions:
\begin{eqnarray*}
   &&\int_{\mathcal{J}\times\mathcal{U}}A^{[m]}(\textbf{v};\textbf{u}_{i_1},\ldots,
      \textbf{u}_{i_m})d\textbf{v}=1, \quad m\geq1.
\end{eqnarray*}
We refer to paper \cite{L11}, where examples of the functions $a^{[m]}$ and $A^{[m]}$ are given
in the context of biological systems.

In case of $m=1$ generator (\ref{gen_obs_gener}) has the form $\sum_{i_1=1}^n\Lambda^{[1]}_n(i_1)$
and it describes the free stochastic evolution of entities. The case of $m\geq2$ corresponds to a
system with the $m$-body interaction of entities in the sense accepted in kinetic theory \cite{KL}.
The $m$-body interaction of entities is the distinctive property of biological systems in comparison
with many-particle systems, for example, gases of atoms with a pair interaction potential.

On the space $C_n$ the one-parameter mapping $e^{t\Lambda_n}$ is a bounded $\ast$-weak continuous
semigroup of operators \cite{BanArl}.

\subsection{The evolution equations for observables and states}
The observables of a system of a non-fixed number of entities of various subpopulations are the
sequences $O=(O_{0},O_{1}(\textbf{u}_1),\ldots,O_{n}(\textbf{u}_1,\ldots,\textbf{u}_n),\ldots)$ of
functions $O_{n}(\textbf{u}_1,\ldots,\textbf{u}_n)$ defined on $(\mathcal{J}\times\mathcal{U})^n$
and $O_{0}$ is a real number. The evolution of observables is described by the sequences
$O(t)=(O_{0},O_{1}(t,\textbf{u}_1),\ldots,O_{n}(t,\textbf{u}_1,\ldots,\textbf{u}_n),\ldots)$
of the functions
\begin{eqnarray*}
   &&O_{n}(t)=e^{t\Lambda_n}O_{n}^0, \quad n\geq1,
\end{eqnarray*}
i.e. they are the corresponding solution of the Cauchy problem of the Liouville equation
(or the Kolmogorov forward equation) with initial data $O_{n}^0$.

The average values of observables (mean values of observables) are determined by the following
positive continuous linear functional defined on the space $C_\gamma$:
\begin{eqnarray}\label{averageD}
     &&\hskip-7mm\langle O\rangle(t)=(I,D(0))^{-1}(O(t),D(0))\doteq
        (I,D(0))^{-1}\sum\limits_{n=0}^{\infty}\frac{1}{n!}
        \int_{(\mathcal{J}\times\mathcal{U})^n} d\textbf{u}_1\ldots d\textbf{u}_{n}\,O_{n}(t)\,D_{n}^0,
\end{eqnarray}
where $D(0)=(1,D_{1}^0,\ldots,D_{n}^0,\ldots)$ is a sequence of nonnegative functions
$D_{n}^0$ defined on $(\mathcal{J}\times\mathcal{U})^n$ that describes the states of
a system of a non-fixed number of entities of various subpopulations at initial time
and $(I,D(0))=\sum_{n=0}^{\infty}\frac{1}{n!}\int_{(\mathcal{J}\times\mathcal{U})^n}d\textbf{u}_1\ldots
d\textbf{u}_{n}\,D_{n}^0$ is a normalizing factor (the grand canonical partition function).

Let $L^{1}_{\alpha}=\oplus^{\infty}_{n=0}\alpha^n L^{1}_{n}$ be the space of sequences
$f=(f_0,f_1,\ldots,f_n,\ldots)$ of the integrable functions $f_n(\textbf{u}_1,\ldots,\textbf{u}_n)$
defined on $(\mathcal{J}\times\mathcal{U})^n$, that are symmetric with respect to permutations
of the arguments $\textbf{u}_1,\ldots,\textbf{u}_n$, and equipped with the norm:
\begin{eqnarray*}
   &&\|f\|_{L^{1}_\alpha}=\sum\limits_{n=0}^\infty\alpha^n\|f_n\|_{L^{1}_n}=
       \sum\limits_{n=0}^\infty\alpha^n\sum\limits_{j_1\in\mathcal{J}}\ldots
       \sum\limits_{j_n\in\mathcal{J}}\,\,
       \int_{\mathcal{U}^n}du_1\ldots du_{n}\big|f_n(\textbf{u}_1,\ldots,\textbf{u}_n)\big|,
\end{eqnarray*}
where $\alpha>1$ is a parameter. Then for $D(0)\in L^{1}$ and $O(t)\in C_\gamma$ average value
functional (\ref{averageD}) exists and it determines a duality between observables and states.

As a consequence of the validity for functional (\ref{averageD}) of the following equality:
\begin{eqnarray*}
     &&(I,D(0))^{-1}(O(t),D(0))=(I,D(0))^{-1}(e^{t\Lambda}O(0)\,D(0))=\\
     &&=(I,e^{t\Lambda^\ast}D(0))^{-1}(O(0)\,e^{t\Lambda^\ast}D(0))\equiv(I,D(t))^{-1}(O(0),D(t)),
\end{eqnarray*}
where $e^{t\Lambda^\ast}=\oplus^{\infty}_{n=0}e^{t\Lambda_n^\ast}$ is the adjoint semigroup
of operators with respect to the semigroup $e^{t\Lambda}=\oplus^{\infty}_{n=0}e^{t\Lambda_n}$,
it is possible to describe the evolution within the framework of the evolution of states.
Indeed, the evolution of all possible states, i.e. the sequence
$D(t)=(1,D_{1}(t,\textbf{u}_1),\ldots, D_{n}(t,\textbf{u}_1,\ldots,$ $\textbf{u}_n),\ldots)\in L^{1}$
of the distribution function $D_{n}(t),\, n\geq1$, is determined by the formula:
\begin{eqnarray*}
   &&D_{n}(t)=e^{t\Lambda^\ast_n}D^0_n, \quad n\geq1,
\end{eqnarray*}
where the operator $\Lambda^\ast_n$ is the adjoint operator to operator
(\ref{gen_obs_gener}) and on $L^{1}_{n}$ it is defined as follows:
\begin{eqnarray}\label{gen_state_gener}
 &&\hskip-7mm (\Lambda^\ast_n f_n)(\textbf{u}_1,\ldots,\textbf{u}_n)\doteq
     \sum_{m=1}^M\varepsilon^{m-1}\sum_{i_1\neq\ldots\neq i_m=1}^n
     \big(\int_{\mathcal{J}\times\mathcal{U}}
     A^{[m]}(\textbf{u}_{i_1};\textbf{v},\textbf{u}_{i_2},\ldots,\textbf{u}_{i_m})a^{[m]}(\textbf{v},\\
 &&\hskip-7mm\textbf{u}_{i_2},\ldots,\textbf{u}_{i_m})f_n(\textbf{u}_1,\ldots,
     \textbf{u}_{{i_1}-1},\textbf{v},\textbf{u}_{{i_1}+1},\ldots,\textbf{u}_n)d\textbf{v}-
     a^{[m]}(\textbf{u}_{i_1},\ldots,\textbf{u}_{i_m})f_n(\textbf{u}_1,\ldots,\textbf{u}_n)\big)\nonumber,
\end{eqnarray}
where the functions $A^{[m]}$ and $a^{[m]}$ are defined as above in (\ref{gen_obs_gener}).

The function $D_{n}(t)$ is a solution of the Cauchy problem of the dual Liouville equation
(or the Kolmogorov backward equation).

On the space $L^{1}_{n}$ the one-parameter mapping $e^{t\Lambda^\ast_n}$ is a bounded strong continuous
semigroup of operators \cite{BanArl}.

%%%%%%%%%%%%%%%%%%%%%%%%%%%%%%%%%%%%%%%%%%%%%%%%%%%%%%%%%%%%%%%%%%%%%%%%%%%%%%%%%%%%%%%%%%%%%%%%%%%%%%%%%%%%%%%%%%%%%%%%%%%%%%%%%%%%%%%%

\section{Hierarchies of evolution equations for entities of various subpopulations}
For the description of microscopic behavior of many-entity systems we also introduce the hierarchies
of evolution equations for marginal observables and marginal distribution functions known as the dual
BBGKY hierarchy and the BBGKY hierarchy, respectively \cite{CGP97}. These hierarchies are constructed
as the evolution equations for one more method of the description of observables and states of finitely
many entities.

\subsection{Marginal observables and states of many-entity systems}
An equivalent approach to the description of observables and states of many-entity systems 
is given in terms of marginal observables $B(t)=(B_0,B_{1}(t,\textbf{u}_1),$ 
$\ldots,B_{s}(t,\textbf{u}_1,\ldots,\textbf{u}_s),\ldots)$ and marginal distribution functions 
$F(0)=(1,F_{1}^0(\textbf{u}_1),\ldots,$ $F_{s}^0(\textbf{u}_1,\ldots,\textbf{u}_s),\ldots)$.
Considering formula (\ref{averageD}), marginal observables and marginal distribution functions
are introduced according to the equality:
\begin{eqnarray*}\label{avmar}
   &&\big\langle O\big\rangle(t)=(I,D(0))^{-1}(O(t),D(0))=(B(t),F(0)),
\end{eqnarray*}
where $(I,D(0))$ is a normalizing factor defined as above. If $F(0)\in L^{1}_{\alpha}$ and $B(0)\in C_\gamma$,
then at $t\in \mathbb{R}$ the functional $(B(t),F(0))$ exists under the condition that: $\gamma>\alpha^{-1}$.

Thus, the relationship of marginal distribution functions $F(0)=(1,F_{1}^0,\ldots,F_{s}^0,\ldots)$
and the distribution functions $D(0)=(1,D_{1}^0,\ldots,D_{n}^0,\ldots)$ is determined by the formula:
\begin{eqnarray*}\label{ms}
      &&\hskip-5mm F_{s}^0(\textbf{u}_1,\ldots,\textbf{u}_s)\doteq(I,D(0))^{-1}\sum\limits_{n=0}^{\infty}\frac{1}{n!}\,
          \int_{(\mathcal{J}\times\mathcal{U})^n}d\textbf{u}_{s+1}\ldots
          d\textbf{u}_{s+n}\,D_{s+n}^0(\textbf{u}_1,\ldots,\textbf{u}_{s+n}), \quad s\geq 1,
\end{eqnarray*}
and, respectively, the marginal observables are determined in terms of the observables as follows:
\begin{eqnarray*}\label{mo}
      &&\hskip-5mm B_{s}(t,\textbf{u}_1,\ldots,\textbf{u}_s)\doteq\sum_{n=0}^s\,
            \frac{(-1)^n}{n!}\sum_{j_1\neq\ldots\neq j_{n}=1}^s
            O_{s-n}\big(t,(\textbf{u}_1,\ldots,\textbf{u}_s)\setminus(\textbf{u}_{j_1},\ldots,\textbf{u}_{j_{n}})\big),
            \quad s\geq 1.\\
\end{eqnarray*}

Two equivalent approaches to the description of the evolution of many interacting entities 
are the consequence of the validity of the following equality for the functional of mean values 
of marginal observables:
\begin{eqnarray*}
   &&(B(t),F(0))=(B(0),F(t)),
\end{eqnarray*}
where $B(0)=(1,B_{1}^0,\ldots,B_{s}^0,\ldots)$ is a sequence of marginal observables at initial moment.

We remark that the evolution of many-entity systems is usually described
within the framework of the evolution of states by the sequence
$F(t)=(1,F_{1}(t,\textbf{u}_1),$ $\ldots,F_{s}(t,\textbf{u}_1,$ $\ldots,\textbf{u}_s),\ldots)$
of the marginal distribution functions $F_s(t,\textbf{u}_1,\ldots,\textbf{u}_s)$ governed by
the BBGKY hierarchy for entities \cite{L11}:
\begin{eqnarray*}
   &&\hskip-7mm \frac{\partial}{\partial t}F_s(t,\textbf{u}_1,\ldots,\textbf{u}_s)=\Lambda^{\ast}_s
      F_{s}(t,\textbf{u}_1,\ldots,\textbf{u}_s)+\sum\limits_{k=1}^{s}\frac{1}{k!}
      \sum\limits_{i_1\neq\ldots\neq i_{k}=1}^{s}\,\sum\limits_{n=1}^{M-k}\frac{\varepsilon^{k+n-1}}{n!}\times\\
   &&\hskip-7mm\int_{(\mathcal{J}\times\mathcal{U})^n}d\textbf{u}_{s+1}\ldots d\textbf{u}_{s+n}
      \sum\limits_{j_1\neq\ldots\neq j_{k+n}\in(i_1,\ldots,i_{k},s+1,\ldots,s+n)}
      \Lambda^{*[k+n]}(j_1,\ldots,j_{k+n})F_{s+n}(t,\textbf{u}_1,\ldots,\textbf{u}_{s+n}),\\ \\
   &&\hskip-7mm s\geq1,
\end{eqnarray*}
where on $L^{1}_{n}$ the adjoint Liouville operator $\Lambda^{\ast}_s$ is defined by formula 
(\ref{gen_state_gener}) and we used notations accepted above.

\subsection{The dual BBGKY hierarchy for entities of various subpopulations}
The evolution of a non-fixed number of entities of various subpopulations within the framework of
marginal observables is described by the following Cauchy problem of the dual BBGKY hierarchy \cite{BG}:
\begin{eqnarray}\label{dh}
   &&\hskip-5mm \frac{d}{dt}B_{s}(t,\textbf{u}_1,\ldots,\textbf{u}_s)=
       \Lambda_s B_{s}(t,\textbf{u}_1,\ldots,\textbf{u}_s)+
       \sum\limits_{n=1}^{s}\frac{1}{n!}\sum\limits_{k=n+1}^s \frac{1}{(k-n)!}\times\\
   &&\hskip-5mm \times\sum_{j_1\neq\ldots\neq j_{k}=1}^s\varepsilon^{k-1}\Lambda^{[k]}(j_1,\ldots,
       j_{k})\sum_{i_1\neq\ldots\neq i_{n}\in(j_1,\ldots,j_{k})}B_{s-n}(t,(\textbf{u}_1,\ldots,
       \textbf{u}_s)\setminus(\textbf{u}_{i_1},\ldots,\textbf{u}_{i_{n}})),\nonumber\\
       \nonumber\\
   \label{dhi}
   &&\hskip-5mm B_s(t,\textbf{u}_1,\ldots,\textbf{u}_s)\mid_{t=0}=
       B_{s}^{0,\varepsilon}(\textbf{u}_1,\ldots,\textbf{u}_s), \quad s\geq 1,
\end{eqnarray}
where the operators $\Lambda_s$ and $\Lambda^{[k]}$ are defined by formulas (\ref{gen_obs_gener}) and
the functions $B_{s}^{0,\varepsilon},\,s\geq 1,$ are a scaled initial data.

The simplest examples of recurrence evolution equations (\ref{dh}) have the form
\begin{eqnarray*}
   &&\hskip-5mm\frac{d}{dt}B_{1}(t,\textbf{u}_1)=\Lambda^{[1]}(1)B_{1}(t,\textbf{u}_1),\\
   &&\hskip-5mm\frac{d}{dt}B_{2}(t,\textbf{u}_1,\textbf{u}_2)=
       (\sum_{i=1}^2\Lambda^{[1]}(i)+\varepsilon\Lambda^{[2]}(1,2))B_{2}(t)
       +\varepsilon(\Lambda^{[2]}(1,2)B_{1}(t,\textbf{u}_1)+\Lambda^{[2]}(2,1)B_{1}(t,\textbf{u}_2)).
\end{eqnarray*}

The solution $B(t)=(B_{0},B_{1}(t,\textbf{u}_1),\ldots,B_{s}(t,\textbf{u}_1,\ldots,$ $\textbf{u}_s),\ldots)$
of the Cauchy problem of recurrence evolution equations (\ref{dh}),(\ref{dhi}) is given by the following
expansions:
\begin{eqnarray}\label{sdh}
   &&\hskip-5mm B_{s}(t,\textbf{u}_1,\ldots,\textbf{u}_s)=
      \sum_{n=0}^s\,\frac{1}{n!}\sum_{j_1\neq\ldots\neq j_{n}=1}^s\mathfrak{A}_{1+n}(t,\{Y\setminus Z\},\,Z)\,
      B_{s-n}^{0,\varepsilon}(\textbf{u}_1,\ldots,\\
   &&\hskip+5mm\textbf{u}_{j_1-1},\textbf{u}_{j_1+1},\ldots,\textbf{u}_{j_n-1},
      \textbf{u}_{j_n+1},\ldots,\textbf{u}_s), \quad s\geq 1,\nonumber
\end{eqnarray}
where the $(1+n)th$-order cumulant of the semigroups $\{e^{t\Lambda_{k}}\}_{t\in\mathbb{R}},\, k\geq1,$
is determined by the formula \cite{BG}:
\begin{eqnarray}\label{cumulantd}
   &&\hskip-5mm\mathfrak{A}_{1+n}(t,\{Y\setminus Z\},\,Z)\doteq
       \sum\limits_{\mathrm{P}:\,(\{Y\setminus Z\},\,Z)={\bigcup}_i Z_i}
       (-1)^{\mathrm{|P|}-1}({\mathrm{|P|}-1})!\prod_{Z_i\subset \mathrm{P}}e^{t\Lambda_{|\theta(Z_i)|}},
\end{eqnarray}
the sets of indexes are denoted by $Y\equiv(1,\ldots,s)$, $Z\equiv(j_1,\ldots,j_{n})\subset Y$, the set
$\{Y\setminus Z\}$ consists from one element $Y\setminus Z=(1,\ldots,j_1-1,j_1+1,\ldots,j_n-1,j_n+1,\ldots,s)$
and the mapping $\theta(\cdot)$ is the declusterization operator defined as follows:
$\theta(\{Y\setminus Z\},\,Z)=Y$.

The simplest examples of marginal observables (\ref{sdh}) are given by the following expansions:
\begin{eqnarray*}
   &&B_{1}(t,\textbf{u}_1)=\mathfrak{A}_{1}(t,1)B_{1}^{0,\varepsilon}(\textbf{u}_1),\\
   &&B_{2}(t,\textbf{u}_1,\textbf{u}_1)=\mathfrak{A}_{1}(t,\{1,2\})
      B_{2}^{0,\varepsilon}(\textbf{u}_1,\textbf{u}_2)+
      \mathfrak{A}_{2}(t,1,2)(B_{1}^{0,\varepsilon}(\textbf{u}_1)+B_{1}^{0,\varepsilon}(\textbf{u}_1)),
\end{eqnarray*}
where first and second order cumulants (\ref{cumulantd}) are determined by the corresponding formulas:
\begin{eqnarray*}
   &&\mathfrak{A}_{1}(t,\{1,2\})=e^{t\Lambda_{2}},\\
   &&\mathfrak{A}_{2}(t,1,2)=e^{t\Lambda_{2}}-e^{t\Lambda^{[1]}(1)}e^{t\Lambda^{[1]}(2)}.
\end{eqnarray*}

For initial data $B(0)=(B_{0},B_{1}^{0,\varepsilon},\ldots,B_{s}^{0,\varepsilon},\ldots)\in\mathcal{C}_{\gamma}$
the sequence $B(t)$ of marginal observables given by expansions (\ref{sdh}) is a classical solution
of the Cauchy problem of the dual BBGKY hierarchy for entities (\ref{dh}),(\ref{dhi}).

%%%%%%%%%%%%%%%%%%%%%%%%%%%%%%%%%%%%%%%%%%%%%%%%%%%%%%%%%%%%%%%%%%%%%%%%%%%%%%%%%%%%%%%%%%%%%%%%%%%%%%%%%%%%%%%%%%%%%%%%%%%%%%%%%%%

\section{The kinetic evolution in terms of marginal observables of entities}
To consider mesoscopic properties of interacting entities we develop an approach to the
description of the kinetic evolution within the framework of the evolution equations for
marginal observables. For this purpose we construct the mean field asymptotics \cite{Sp80}
of a solution of the Cauchy problem of the dual BBGKY hierarchy for entities (see also
\cite{G11},\cite{G}).

\subsection{The mean field limit of a solution of the dual BBGKY hierarchy for entities}
We restrict ourself by the case of $M=2$ subpopulations to simplify the cumbersome formulas
and consider the mean field scaling limit of nonperturbative solution (\ref{sdh}) of the Cauchy
problem of the dual BBGKY hierarchy for entities (\ref{dh}),(\ref{dhi}).

Let for initial data $B_{s}^{0,\varepsilon}\in\mathcal{C}_s$ there exists the limit function
$b_{s}^0\in\mathcal{C}_s$
\begin{eqnarray*}\label{asumdin}
    &&\mathrm{w^{\ast}-}\lim\limits_{\varepsilon\rightarrow 0}\big(\varepsilon^{-s}
         B_{s}^{0,\varepsilon}-b_{s}^0\big)=0,
\end{eqnarray*}
then for arbitrary finite time interval there exists the mean field limit of solution (\ref{sdh})
of the Cauchy problem of the dual BBGKY hierarchy for entities  (\ref{dh}),(\ref{dhi}) in the sense 
of the $\ast$-weak convergence of the space $\mathcal{C}_s$
\begin{eqnarray}\label{asymt}
   && \mathrm{w^{\ast}-}\lim\limits_{\varepsilon\rightarrow 0} \big(\varepsilon^{-s}B_{s}(t)-b_{s}(t)\big)=0,
\end{eqnarray}
and it is determined by the expansion:
\begin{eqnarray}\label{Iterd}
   &&\hskip-8mm b_{s}(t,\textbf{u}_1,\ldots,\textbf{u}_s)=\\
   &&\hskip-8mm=\sum\limits_{n=0}^{s-1}\,\int_0^tdt_{1}\ldots\int_0^{t_{n-1}}dt_{n}
      \,e^{(t-t_{1})\sum\limits_{k_{1}=1}^{s}\Lambda^{[1]}(k_{1})}\sum\limits_{i_{1}\neq j_{1}=1}^{s}
      \Lambda^{[2]}(i_{1},j_{1})e^{(t_{1}-t_{2})\sum\limits_{l_{1}=1,l_{1}\neq j_{1}}^{s}\Lambda^{[1]}(l_{1})}
       \ldots\nonumber\\
   &&\hskip-8mm e^{(t_{n-1}-t_{n})\hskip-1mm\sum\limits^{s}_{\mbox{\scriptsize $\begin{array}{c}k_{n}=1,\\
      k_{n}\neq (j_{1},\ldots,j_{n-1}))\end{array}$}}\hskip-1mm\Lambda^{[1]}(k_{n})}
            \hskip-2mm\sum\limits^{s}_{\mbox{\scriptsize $\begin{array}{c}i_{n}\neq j_{n}=1,\\
      i_{n},j_{n}\neq (j_{1},\ldots,j_{n-1})\end{array}$}}\hskip-1mm\Lambda^{[2]}(i_{n},j_{n})
      e^{t_{n}\hskip-1mm\sum\limits^{s}_{\mbox{\scriptsize $\begin{array}{c}l_{n}=1,\\
      l_{n}\neq (j_{1},\ldots,j_{n}))\end{array}$}}\hskip-1mm\Lambda^{[1]}(l_{n})}
      b_{s-n}^0((\textbf{u}_1,\nonumber\\
   &&\hskip-8mm\ldots,\textbf{u}_s)\setminus (\textbf{u}_{j_{1}},\ldots,\textbf{u}_{j_{n}})).\nonumber
\end{eqnarray}

The proof of this statement is based on formulas for cumulants of asymptotically perturbed semigroups
of operators (\ref{cumulantd}).

If $b^0\in\mathcal{C}_{\gamma}$, then the sequence $b(t)=(b_0,b_1(t),\ldots,b_{s}(t),\ldots)$
of limit marginal observables (\ref{Iterd}) is generalized global in time solution of the Cauchy
problem of the dual Vlasov hierarchy for entities:
\begin{eqnarray}\label{vdh}
   &&\hskip-5mm \frac{\partial}{\partial t}b_{s}(t)=
     \sum\limits_{j=1}^{s}\Lambda^{[1]}(j)\,b_{s}(t)+
     \sum_{j_1\neq j_{2}=1}^s\Lambda^{[2]}(j_1,j_{2})\,b_{s-1}(t,\textbf{u}_1,\ldots,\textbf{u}_{j_{2}-1},
       \textbf{u}_{j_{2}+1},\ldots,\textbf{u}_s),\\ \nonumber\\
  \label{vdhi}
   &&\hskip-5mm b_{s}(t,\textbf{u}_1,\ldots,\textbf{u}_s)\mid_{t=0}=b_{s}^0(\textbf{u}_1,\ldots,\textbf{u}_s),
       \quad s\geq1,
\end{eqnarray}
where in recurrence evolution equations (\ref{vdh}) the operators $\Lambda^{[1]}(j)$ and
$\Lambda^{[2]}(j_1,j_{2})$ are defined by formula (\ref{gen_obs_gener}).

\subsection{The relationships of kinetic equations for marginal observables and states}
We consider initial states of statistically independent entities specified by a one-particle
marginal distribution function, namely $f^{(c)}\equiv(1,f_1^0(\textbf{u}_1),\ldots,
{\prod\limits}_{i=1}^{s}f_{1}^0(\textbf{u}_i),\ldots)$. Such states are intrinsic
for the kinetic description of many-entity systems \cite{CGP97},\cite{Sp80}.

If $b(t)\in\mathcal{C}_{\gamma}$ and $f_1^0\in L^{1}(\mathcal{J}\times\mathcal{U})$,
then under the condition that: $\|f_1^0\|_{L^{1}(\mathcal{J}\times\mathcal{U})}<\gamma$,
there exists the mean field scaling limit of the mean value functional of marginal 
observables and it is determined by the following series expansion:
\begin{eqnarray*}
   &&\big\langle b(t)\big|f^{(c)}\big\rangle=\sum\limits_{s=0}^{\infty}\,\frac{1}{s!}\,
      \int_{(\mathcal{J}\times\mathcal{U})^s}d\textbf{u}_{1}\ldots d\textbf{u}_{s}
      \,b_{s}(t,\textbf{u}_1,\ldots,\textbf{u}_s)\prod\limits_{i=1}^{s} f_1^0(\textbf{u}_i).
\end{eqnarray*}

Then for the mean value functionals of the limit additive-type marginal observables
the following representation is true:
\begin{eqnarray}\label{avmar-2}
   &&\hskip-7mm\big\langle b^{(1)}(t)\big|f^{(c)}\big\rangle=\sum\limits_{s=0}^{\infty}\,\frac{1}{s!}\,
       \int_{(\mathcal{J}\times\mathcal{U})^s}d\textbf{u}_{1}\ldots d\textbf{u}_{s}
       \,b_{s}^{(1)}(t,\textbf{u}_1,\ldots,\textbf{u}_s)\prod \limits_{i=1}^{s} f_{1}^0(\textbf{u}_i)=\\
   &&=\int_{(\mathcal{J}\times\mathcal{U})}d\textbf{u}_{1}\,
       b_{1}^{(1)}(0,\textbf{u}_{1})f_{1}(t,\textbf{u}_{1}).\nonumber
\end{eqnarray}
In equality (\ref{avmar-2}) the function $b_{s}^{(1)}(t)$ is given by a special case of
expansion (\ref{Iterd}), namely
\begin{eqnarray*}\label{itvad}
   &&\hskip-5mm b_{s}^{(1)}(t,\textbf{u}_1,\ldots,\textbf{u}_s)=\\
   &&\hskip-5mm =\int_0^t dt_{1}\ldots\int_0^{t_{s-2}}dt_{s-1}
       \,e^{(t-t_{1})\sum\limits_{k_{1}=1}^{s}\Lambda^{[1]}(k_{1})}
       \sum\limits_{i_{1}\neq j_{1}=1}^{s}\Lambda^{[2]}(i_{1},j_{1})\,
       e^{(t_{1}-t_{2})\sum\limits_{l_{1}=1,\,\,l_{1}\neq j_{1}}^{s}\Lambda^{[1]}(l_{1})}\\
   &&\hskip-5mm\ldots\,e^{(t_{s-2}-t_{s-1})\sum\limits_{k_{s-1}=1,\,\,k_{s-1}\neq (j_{1},\ldots,j_{s-2})}^{s}
       \Lambda^{[1]}(k_{s-1})}\sum\limits^{s}_{\mbox{\scriptsize $\begin{array}{c}i_{s-1}\neq j_{s-1}=1,\\
       i_{s-1},j_{s-1}\neq (j_{1},\ldots,j_{s-2})\end{array}$}}\Lambda^{[2]}(i_{s-1},j_{s-1})\\
   &&\hskip-5mm\times e^{t_{s-1}\sum\limits_{l_{s-1}=1,\,\,l_{s-1}\neq (j_{1},\ldots,j_{s-1})}^{s}
       \Lambda^{[1]}(l_{s-1})}\,
       b_{1}^{(1)}(0,(\textbf{u}_1,\ldots,\textbf{u}_s)\setminus(\textbf{u}_{j_{1}},\ldots,\textbf{u}_{j_{s-1}})),
       \quad s\geq1,
\end{eqnarray*}
and the limit one-particle distribution function $f_{1}(t)$ is represented by the series expansion:
\begin{eqnarray}\label{viter}
   &&\hskip-9mm f_{1}(t,\textbf{u}_1)=\\
   &&\hskip-9mm = \sum\limits_{n=0}^{\infty}\int_0^tdt_{1}\ldots\int_0^{t_{n-1}}dt_{n}
      \int_{(\mathcal{J}\times\mathcal{U})^n}d \textbf{u}_{2}\ldots d \textbf{u}_{n+1}\,
      e^{(t-t_{1})\Lambda^{\ast[1]}(1)}\Lambda^{\ast[2]}(1,2)
      \prod\limits_{j_1=1}^{2}e^{(t_{1}-t_{2})\Lambda^{\ast[1]}(j_1)}\ldots\nonumber\\
   &&\hskip-9mm\prod\limits_{j_{n-1}=1}^{n}e^{(t_{n-1}-t_{n})\Lambda^{\ast[1]}(j_{n-1})}
      \sum\limits_{i_{n}=1}^{n}\Lambda^{\ast[2]}(i_{n},n+1)
      \prod\limits_{j_n=1}^{n+1}e^{t_{n}\Lambda^{\ast[1]}(j_{n})}
      \prod\limits_{i=1}^{n+1}f_{1}^0(\textbf{u}_i),\nonumber
\end{eqnarray}
where the operator $\Lambda^{\ast[2]}$ is defined as a particular case of the  
operators $\Lambda^{\ast[m]},\,m\geq1$:
\begin{eqnarray*}
 &&\hskip-8mm \Lambda^{\ast[m]}(i_1,\ldots,i_m)f_n(\textbf{u}_1,\ldots,\textbf{u}_n)=
     \big(\int_{\mathcal{J}\times\mathcal{U}}
     A^{[m]}(\textbf{u}_{i_1};\textbf{v},\textbf{u}_{i_2},\ldots,\textbf{u}_{i_m})
     a^{[m]}(\textbf{v},\textbf{u}_{i_2},\ldots,\textbf{u}_{i_m})\times\nonumber\\
 &&\times f_n(\textbf{u}_1,\ldots,\textbf{u}_{{i_1}-1},\textbf{v},
     \textbf{u}_{{i_1}+1},\ldots,\textbf{u}_n)d\textbf{v}-
     a^{[m]}(\textbf{u}_{i_1},\ldots,\textbf{u}_{i_m})f_n(\textbf{u}_1,\ldots,\textbf{u}_n)\big)\nonumber.
\end{eqnarray*}

For initial data $f_{1}^0\in L^{1}(\mathcal{J}\times\mathcal{U})$ limit marginal distribution
function (\ref{viter}) is a strong solution of the Cauchy problem of the Vlasov kinetic equation
for entities \cite{L11}:
\begin{eqnarray}
  \label{Vlasov1}
    &&\frac{\partial}{\partial t}f_{1}(t,\textbf{u}_1)= \Lambda^{\ast[1]}(1)f_{1}(t,\textbf{u}_1)+
       \int_{\mathcal{J}\times\mathcal{U}}d\textbf{u}_{2}
       \Lambda^{\ast[2]}(1,2)f_{1}(t,\textbf{u}_1)f_{1}(t,\textbf{u}_2),\\ \nonumber\\
  \label{Vlasov2}
    &&f_{1}(t,\textbf{u}_1)|_{t=0}=f_1^0(\textbf{u}_1), \nonumber
\end{eqnarray}
or in case of general interactions of entities (\ref{gen_obs_gener}) the Vlasov kinetic equation
takes the form
\begin{eqnarray*}
     &&\hskip-8mm \frac{\partial}{\partial t}f_{1}(t,\textbf{u}_1)=\Lambda^{\ast[1]}(1)f_{1}(t,\textbf{u}_1)+\\
     &&\hskip-8mm + \sum\limits_{n=1}^{M-1}\frac{1}{n!}
        \int_{(\mathcal{J}\times\mathcal{U})^n}d\textbf{u}_{2}\ldots d\textbf{u}_{n+1}
        \sum\limits_{j_1\neq\ldots\neq j_{n+1}\in(1,\ldots,n+1)}
        \Lambda^{\ast[n+1]}(j_1,\ldots,j_{n+1})\prod\limits_{i=1}^{n+1}f_{1}(t,\textbf{u}_i).
\end{eqnarray*}

Correspondingly, in the Heisenberg picture of the evolution of entities a chaos property 
fulfils. It follows from the equality for the mean value functionals of the limit
$k$-ary  marginal observables, i.e. the sequences
$b^{(k)}(0)=(0,\ldots,0,b_{k}^{(k)}(0,\textbf{u}_1,\ldots,\textbf{u}_k),0,\ldots)$,
\begin{eqnarray}\label{dchaos}
    &&\hskip-7mm\big\langle b^{(k)}(t)\big|f^{(c)}\big\rangle=\sum\limits_{s=0}^{\infty}\,\frac{1}{s!}\,
       \int_{(\mathcal{J}\times\mathcal{U})^s}d\textbf{u}_{1}\ldots d\textbf{u}_{s}
       \,b_{s}^{(k)}(t,\textbf{u}_1,\ldots,\textbf{u}_s) \prod \limits_{i=1}^{s} f_1^0(\textbf{u}_i)=\\
    &&\hskip-7mm=\frac{1}{k!}\int_{(\mathcal{J}\times\mathcal{U})^k}d\textbf{u}_{1}\ldots d\textbf{u}_{k}
       \,b_{k}^{(k)}(0,\textbf{u}_1,\ldots,\textbf{u}_k)
       \prod\limits_{i=1}^{k} f_{1}(t,\textbf{u}_i),\quad k\geq2,\nonumber
\end{eqnarray}
where the limit one-particle marginal distribution function $f_{1}(t,\textbf{u}_i)$ is determined
by series expansion (\ref{viter}).

Thus, an equivalent approach to the description of the kinetic evolution of entities in terms of
the Vlasov-type kinetic equation (\ref{Vlasov1}) is given by the dual Vlasov hierarchy (\ref{vdh}) for
the additive-type marginal observables. In case of the $k$-ary marginal observables the evolution
governed by the dual Vlasov hierarchy (\ref{vdh}) means the propagation of initial chaos (\ref{dchaos})
in terms of the $k$-particle marginal distribution functions.

%%%%%%%%%%%%%%%%%%%%%%%%%%%%%%%%%%%%%%%%%%%%%%%%%%%%%%%%%%%%%%%%%%%%%%%%%%%%%%%%%%%%%%%%%%%%%%%%%%%%%%%%%%%%%%%%%%

\section{Conclusion and outlook}
We developed a new approach to the description of kinetic evolution of large number
of interacting constituents (entities) of mathematical biology within the framework
of the evolution of marginal observables of these systems. Such representation of the
kinetic evolution seems in fact the direct mathematically fully consistent formulation
modeling collective behavior of biological systems, since the notion of state is more
subtle and implicit characteristic of living entities.

A mean field scaling asymptotics of non-perturbative solution (\ref{sdh}) of the dual
BBGKY hierarchy (\ref{dh}) for marginal observables is constructed. We established that
the limit additive-type marginal observables governed by the dual Vlasov hierarchy
(\ref{vdh}) gives an equivalent approach to the description of the kinetic evolution
of many entities in terms of a one-particle distribution function governed by the
Vlasov kinetic equation (\ref{Vlasov1}). Moreover, the kinetic evolution of the $k$-ary
marginal observables governed by the dual Vlasov hierarchy means the property of the
propagation of initial chaos (\ref{dchaos}) for the $k$-particle marginal distribution
functions within the framework of the evolution of states.

One of the advantages of suggested approach in comparison with the conventional approach
of the kinetic theory \cite{CIP},\cite{CGP97},\cite{SR12} is the possibility to construct
kinetic equations in scaling limits in the presence of correlations at initial time which
can characterize the analogs of condensed states of systems of statistical mechanics for
interacting entities or soft active matter \cite{GTsm}.

We note that the developed approach is also related to the problem of a rigorous derivation
of the non-Markovian kinetic-type equations from underlaying many-entity dynamics \cite{G}
which make it possible to describe the memory effects of collective dynamics of entities
modeling systems in mathematical biology.

\subsection*{Acknowledgments}
This work was partially supported by the FP7-People-2011-IRSES project number 295164.
%% References

\addcontentsline{toc}{section}{References}
{\small
\renewcommand{\refname}{References}

\end{document}